\documentclass[preprint,journal]{vgtc}       

\ifpdf
  \pdfoutput=1\relax                   
  \usepackage{graphicx}                
  \DeclareGraphicsExtensions{.pdf,.png,.jpg,.jpeg} 
\else
  \usepackage{graphicx}                
  \DeclareGraphicsExtensions{.eps}     
\fi%

\graphicspath{{figures/}{pictures/}{images/}{./}} 

\usepackage{microtype}                 
\PassOptionsToPackage{warn}{textcomp}  
\usepackage{textcomp}                  
\usepackage{mathptmx}                  
\usepackage{times}                     
\usepackage{cite}                      
\usepackage{tabu}                      
\usepackage{booktabs}                  
\usepackage{hyperref}
\usepackage{amsmath}
\usepackage{multirow}
\usepackage{threeparttable}
\usepackage{amsmath} 
\usepackage{amssymb}
\usepackage{url}

\ieeedoi{10.1109/TVCG.2021.3114789}

\onlineid{1107}

\vgtccategory{Research}
\vgtcpapertype{application/design study}

\title{E-ffective: A Visual Analytic System for Exploring the Emotion and Effectiveness of Inspirational Speeches}

\author{Kevin Maher, Zeyuan Huang, Jiancheng Song, Xiaoming Deng, \textit{Member, IEEE}, Yu-Kun Lai, \textit{Member, IEEE}, \\Cuixia Ma, Hao Wang, Yong-Jin Liu, \textit{Senior Member, IEEE}, Hongan Wang, \textit{Member, IEEE}}
\authorfooter{
\item
 K. Maher is with Institute of Software, Chinese Academy of Sciences and Tsinghua University. E-mail: kevintmaher@gmail.com.

\item
 Z.Y. Huang, J.C. Song, X.M. Deng, C.X. Ma, and H.A. Wang are with State Key Laboratory of Computer Science and Beijing Key Lab of Human-Computer Interaction, Institute of Software, Chinese Academy of Sciences and University of Chinese Academy of Sciences. E-mail: \{zeyuan2020, jiancheng2019, xiaoming, cuixia, hongan\}@iscas.ac.cn.
 
 \item
 Y.-K. Lai is with Cardiff University. E-mail: LaiY4@cardiff.ac.uk.
 
 \item
 H. Wang is with Alibaba Group. E-mail: cashenry@126.com.
 
 \item
 Y.-J. Liu is with Tsinghua University. E-mail: liuyongjin@tsinghua.edu.cn.
 
 \item
 K. Maher and Z.Y. Huang contribute equally to this work. C.X. Ma, H. Wang, Y.-J. Liu are corresponding authors. 
}

\shortauthortitle{Maher and Huang \MakeLowercase{\textit{et al.}}: E-ffective: A Visual Analytic System for Exploring the Emotion and Effectiveness of Inspirational Speeches}

\abstract{What makes speeches effective has long been a subject for debate, and until today there is broad controversy among public speaking experts about what factors make a speech effective as well as the roles of these factors in speeches. Moreover, there is a lack of quantitative analysis methods to help understand effective speaking strategies. In this paper, we propose E-ffective, a visual analytic system allowing speaking experts and novices to analyze both the role of speech factors and their contribution in effective speeches. From interviews with domain experts and investigating existing literature, we 
identified important factors to consider in inspirational speeches. We obtained the generated factors from multi-modal data that were then related to effectiveness data. Our system supports rapid understanding of critical factors in inspirational speeches, including the influence of emotions by means of novel visualization methods and interaction. Two novel visualizations include E-spiral (that shows the emotional shifts in speeches in a visually compact way) and E-script (that connects speech content with key speech delivery information). In our evaluation we studied the influence of our system on experts' domain knowledge about speech factors. We further studied the usability of the system by speaking novices and experts on assisting analysis of inspirational speech effectiveness. %
} 


\CCScatlist{ 
 \CCScat{K.6.1}{Management of Computing and Information Systems}%
{Project and People Management}{Life Cycle};
 \CCScat{K.7.m}{The Computing Profession}{Miscellaneous}{Ethics}
}

\vgtcinsertpkg

\begin{document}


\firstsection{Introduction}

\maketitle

Effectiveness in speeches has long been a controversial subject. Aristotle began his work Rhetoric by criticizing contemporary experts in the field “the framers of the current treatises on rhetoric have constructed but a small portion of that art” \cite{aristotle2009rhys}. While Aristotle focused on persuasive speaking, the effectiveness of other kinds of public speaking remains to be understood on a large scale. Today organizations and public speaking experts train and coach people to improve their public speaking ability in various contexts. However, many principles of public speaking are yet to be agreed upon, even in specific contexts such as speech contests. In certain contexts, a clear metric of success can be given, and with automated annotation of speech factors, in theory a system could allow users to systematically understand the influence of many factors on the effectiveness of speeches at a large scale.

Currently, while there is support from academic literature that different speaking strategies have important impacts in a variety of fields, there is controversy among public speaking experts about the significance and role of different strategies. Recently published works in analytical systems have focused on describing speaking strategies of high level speeches. However, there is a lack of analytical systems that support evaluation of the effectiveness of speaking strategies. Additionally, existing works cannot determine what strategies are effective and what are not, since they lack a clear metric of effectiveness. 

We propose E-ffective, a visual analytic system for speech experts and novices to evaluate the effectiveness of speaking strategies. Our system supports analysis of speech factors that takes into account the relative success of speeches. The system was designed to provide intuitive and understandable visualizations for speaking experts and novices to understand factors in an inspirational speech contest. In our work we seek to investigate speech effectiveness in a user centered approach. The contributions of E-ffective include:
 
 \begin{itemize}

\item \textbf{A visual analysis system supporting analysis of speech effectiveness.} The system allows public speaking experts and novices to explore speech effectiveness in a systematic way that fits with mainstream theory on public speaking. The system was designed to support understanding given moderate visualization literacy of speaking experts and novices. Verification of the understanding of effective speaking patterns as well as the utility of our system was tested in user trials.

\item \textbf{Novel techniques of expressive visualization.} New visualization methods were developed to assist users to further understand critical factor data. These include a spiral-like graph for indicating emotional fluctuations, as well as a method for displaying key information in speech texts. 

\item \textbf{Analytical insights based on an inspirational speech dataset.} We utilize algorithms and visualization methods to identify what factors are indicative of effectiveness. From the evaluation, we reveal the ability of the system to assist users to examine their existing hypothesis about speech factors as well as to generate new hypothesis based on large scale data.

\end{itemize}

\begin{figure*}[t]
 \centering 
 \includegraphics[width=\linewidth]{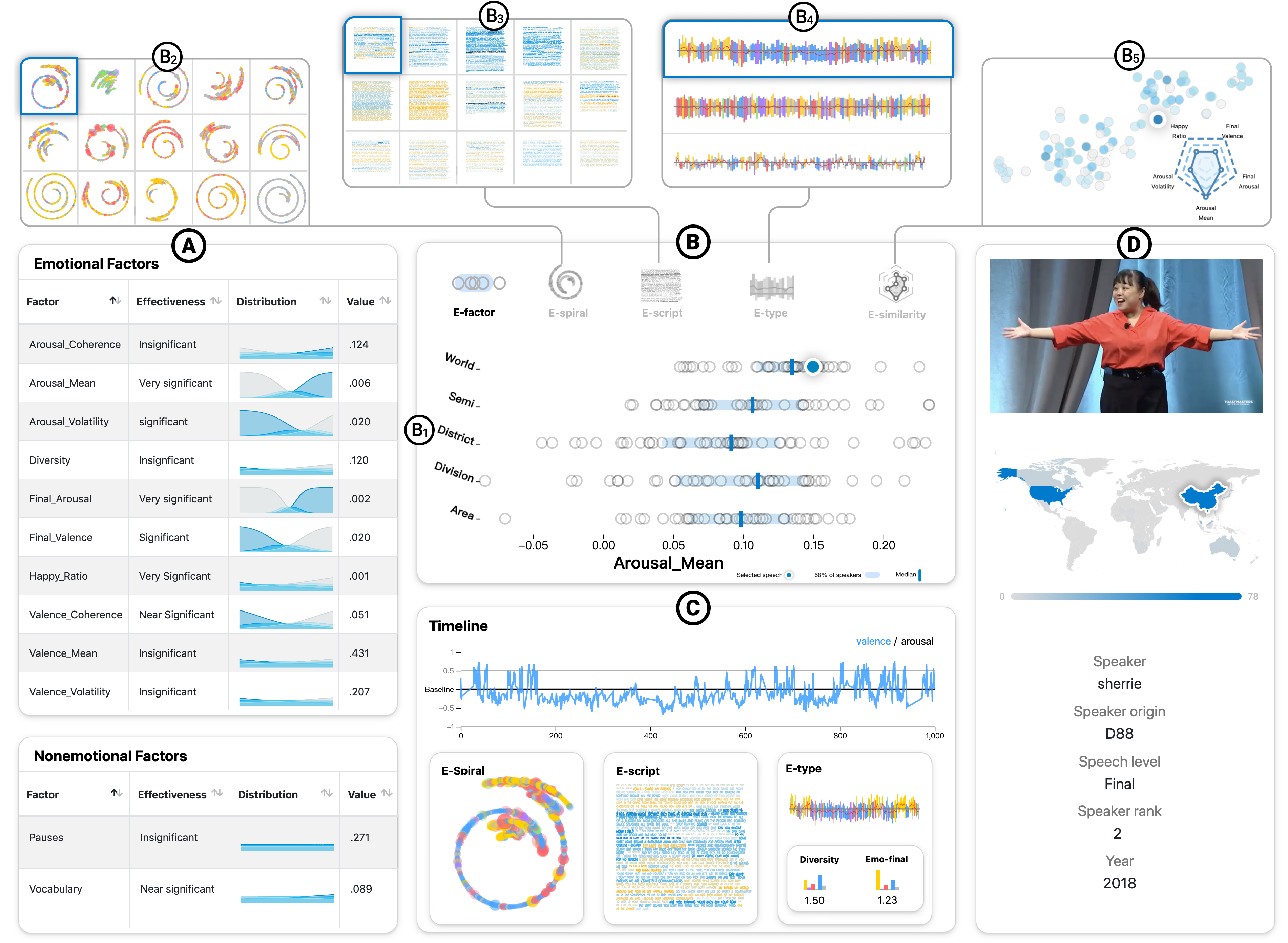}
\setlength{\belowcaptionskip}{-0.5cm}
  \caption{Our E-ffective system supports understanding and exploring the effectiveness of different factors in public speaking in 4 views. (A) The view of factors summarizes different factors extracted from the video and p-values calculated by ordinal regression. (B) The view of all speeches contains 5 sub-views that show the distributions of all speeches. (C) The view of the selected speech supports exploring the visualizations of an individual speech. (D) The view of the selected speech context shows related information about the contest time, place and result (for speech contest videos). }
 \label{fig:MainInterface}
\end{figure*}

\section{Related Work}

\subsection{Effectiveness of Speeches}

There is disagreement from public speaking experts and academics about what techniques and theories support effective speeches. In the $20^{\rm th}$ century various scholars sought to better understand effective communication combining manual annotation of speech metrics and quantitative analysis\cite{barker1968two,lull1940effectiveness,haiman1949experimental}. Speech effectiveness has been measured in a number of ways that include recall-comprehension, attitude change, and the perceived credibility of the speaker\cite{gundersen1976relationships}.

Many of the public speaking principles in common use today originated from these studies, even if they are misapplied.  Possibly the most famous statistic related to communication is Mehrabian’s research \cite{lapakko2007communication} that states ``7\% of communication comes from spoken word, 38\% comes from speech tone and 55\% comes from body language." Lapakko noted the widespread misuse of this research in public speaking textbooks. Still other researchers diametrically opposed applying this research to public speaking, ``the fact is Professor Mehrabian's research had nothing to do with giving speeches ... I have no doubt that the verbal (what you say) must dominate by a wide margin." \cite{yaffe20117} Among public speaking experts, the role of each speech factor is just as disputed as the relative contribution of different factors in effectiveness, as shown by our research survey in Sect. 4.3 of supplementary material. Academic research and speaking experts often have different conclusions. In a book written by a past champion of the World Championship of Public Speaking, the author claimed a general rule is to ``use ... short sentences to express your message."\cite{donovan2014speaker} However, an academic study of educational videos concluded that ``the complexity of the sentences is higher than the ones in the less popular videos."\cite{kravvaris2014speakers} Finding out whether public speaking rules can be generalized to different contexts, such as different audiences or speaking purposes, and evaluating supposed rules of public speaking effectiveness could potentially be achieved by an analytic system with sufficient data. 

Automated analysis of speaking data can allow for understanding of the effectiveness of speeches in different contexts. Recently a few works have dealt with assessing speech effectiveness. In terms of winning and losing results in a speech contest, Fuyuno et al. \cite{Fuyuno2017SemanticSS} focused on studying face rotation angle and speech pause data as correlated to speech contest winners. Ramanarayanan et al. \cite{ramanarayanan2015evaluating} automated scoring of multimodal speech data in terms of multiple dimensions of effectiveness. Kravvaris et al. \cite{kravvaris2014speakers} used YouTube’s automated data extraction in speeches.

Outside of academia, experts in public speaking have written books on how to win speech contests, and claim to understand the most important factors for effectively winning such competitions. Emotion is considered crucial in published books of speech contest winners and led to a world champion giving the definition of a speech at the competition as ``an emotional rollercoaster ride for the speaker and his audience.''~\cite{jhingran2014emote} In their work, experts in these competitions advise candidates how to plan the emotions in their speeches. A different world champion advises: ``Speeches that focus on only one or two emotions risk disengaging the audience. . . the most effective choices are anger, disgust, fear, happiness, love, sadness, and surprise."\cite{donovan2014speaker} Claims by such speech experts about the effectiveness of emotion in speech outcomes have yet to be assessed by quantitative methods. In addition to a survey of literature, in \autoref{section:interviews} we interview speech experts to determine emotional factors considered important for effectiveness.

\subsection{Analysis Systems Supporting the Understanding of Speaking Effectiveness}

To better evaluate different factors in public speaking, visual analytical tools have been developed that allow for large scale understanding. Wu et al. \cite{wu2018multimodal} developed a system that allowed users to understand posture, gesture, and rhetoric in TED videos, allowing users to ``help gain empirical insights into effective presentation delivery." Zeng et al. \cite{zeng2019emoco} developed a system that explores emotional coherence which has a case study where experts help teach how to ``express emotions more effectively." These systems focus on describing different patterns of emotions in speakers considered to be high level. However there is a lack of analytical systems that focus on a clear metric for measuring success, and thus give insight into what differentiates more successful speeches. Claims by speech experts outside of academia as well as predicted patterns made by speech experts in our user survey can be empirically evaluated with our system. In the past,

much work was done by small scale manual annotation to understand speech effectiveness data. Our system extends insights from large scale analysis to public speaking novices and experts.

\subsection{Emotion Visualization}

Visual analysis has been shown to provide valuable insights into emotional speech data. Recently several works have created novel visual analysis techniques aiming for presenting emotional data that provide meaningful insights. New visual forms were developed to show changing group emotion\cite{zeng2020emotioncues} and emotional coherence\cite{zeng2019emoco} in an intuitive way.

As we will later show, emotional data is critical to the effectiveness in the speeches we study. We created visualizations to express several critical metrics in a way designed to be intuitively understandable to a non-expert audience.

\section{Domain-Centered Design}

We adopted a domain-centered design procedure to investigate the comprehension of effective strategies in the domain of public speaking. In this section, we first introduce how we conducted interviews among public speaking experts. Next, we summarize their opinions into overall goals, and the design tasks of the system.

\subsection{In-Depth Interviews} \label{section:interviews}

We conducted initial in-depth interviews with public speaking experts in order to analyze what factors they thought were critical to speech effectiveness, as well as to establish design requirements for an interface that supports their comprehension. We focused our studies on the World Championship of Public Speaking, which in some years has claimed to be the world's most popular speech contest.

The seven experts we interviewed were all professional public speaking trainers that deliver courses for preparing for speech contests. Six of the seven had participated multiple times in the World Championship of Public Speaking. The interview was semi-structured, with all participants asked to list possible critical factors to study. They were also prompted with factors surveyed from literature, to obtain their opinion on the importance and role of the factors on contest effectiveness.
Consistent with literature, all experts thought that the emotion of speakers would have important impact on the outcome of the contest. Among the factors they listed, there are factors that we did not include in our research that are likely to have influence on speech effectiveness, such as gestures and voice pitch. However, in this contest, emotion is viewed as critical, with entire books about how to plan one's emotion~\cite{jhingran2014emote}. We saw that addressing the lack of quantitative methods to evaluate such emotion as a contribution to domain knowledge.

The emotional factors are listed in \autoref{tab:pvalue}. The modality of each factor is indicated by V (voice), F (facial), or T (text). Some of the factors to be studied were determined by a survey of literature. Our later pre-survey in \autoref{sec:user survey} confirmed that experts view these factors as significant in speakers in the competition.  

We also established non-emotional factors that would be critical to assess speech effectiveness. For example, in the interviews, unprompted, three experts suggested that cross-cultural influences would be important for the outcome of the contest, and that the effect of culture on factors in the contest would be prominent. In \autoref{tab:pvalue}, two additional non-emotional factors for comparison, namely pauses and vocabulary, were included that were estimated by the experts to have significant impact on effectiveness.

\subsection{Overall Goals}

 The preliminary interviews and literature review together led to the overall goals to guide our research. 

\textbf{G1: To understand the relation between speech effectiveness and various speech factors.} The relative importance of different speech factors and the role of factors on effectiveness in the contest are critical for users.

\textbf{G2: To understand the spatio-temporal distribution of factors across multiple speeches.} Referenced work showed that experts believed certain factors were more important at different moments of speeches or that the time order of certain factors was important. The geographical distribution was also considered important by some domain experts.

\textbf{G3: To understand the effectiveness of individual speeches in context.} Speech experts in preliminary interviews expressed interest in wanting to understand the patterns of an individual speech. Furthermore how these factors in one speech relate to the factors and effectiveness metrics of all speeches.

\textbf{G4: To compare between speeches on various speech factors.} Observing similarity and differences of speeches can allow effectiveness metrics to be connected with speaking styles. 

\textbf{G5: To understand speaking strategies among speech factors.} As revealed in our literature survey, there are different opinions about how different factors are effectively used. These theories could be evaluated.

\begin{table}[]\small
\centering
  \caption{Emotional and non-emotional factors and the p-values of factors.} 
  \label{tab:pvalue}
\begin{tabular}{ccll}
\hline
Factor                      & Modality                & Type(p-value)                     & Type(p-value)      \\ \hline
\multirow{3}{*}{Average}       & Facial                  & Arousal($0.006^*$)                & Valence(0.431)     \\
                            & Textual                 & Arousal(0.215)                    & Valence(0.088)     \\
                            & Vocal                   & Arousal($0.016^*$)                & Valence($0.017^*$) \\ \hline
\multirow{3}{*}{Volatility} & Facial                  & Arousal($0.020^*$)                & Valence($0.006^*$) \\
                            & Textual                 & Arousal(0.433)                    & Valence(0.438)     \\
                            & Vocal                   & Arousal(0.235)                    & Valence(0.845)     \\ \hline
Diversity                   & Facial                  & \multicolumn{2}{c}{Across Emotion Type(0.120)}         \\ \hline
Final                       & Facial                    & Arousal($0.002^*$)                & Valence($0.020^*$) \\ \hline
Coherence                   & All                     & Arousal(0.124)                   & Valence(0.051) \\ \hline
\multirow{4}{*}{Ratio}      & \multirow{4}{*}{Facial} & Happy($0.001^*$) & Sad(0.0736)        \\
                            &                         & Fear(0.582)                       & Angry(0.292)       \\
                            &                         & Surprise(0.115)                   & Disgust(0.306)     \\
                            &                         & Neutral(0.488)                    &  \multicolumn{1}{c}{-}                  \\ \hline
Pauses                      & Vocal                   & Pauses(0.271)                     & \multicolumn{1}{c}{-}                 \\ \hline
Vocabulary                  & Textual                 & Vocabulary(0.089)                 & \multicolumn{1}{c}{-}                 \\ \hline
\end{tabular}
    \begin{tablenotes}
        \item *: The factor has a significant correlation with speech effectiveness.
    \end{tablenotes}
\end{table}

\subsection{Design Tasks} \label{section:tasks}

\begin{table*}[]\small
 \centering
  \caption{Categorized Design Tasks and Corresponding Goals.} 
  \label{tab:tasks}
\begin{tabular}{llll}
\hline
\textbf{Category}                                & \multicolumn{2}{l}{\textbf{Design Task}}                                   & \textbf{Overall Goals} \\ \hline
\multirow{2}{*}{Visual Data Fusion}              & T1  & To present temporal and geographical distributions of data.                                  &   G1-G5                     \\
                                                 & T2  & To display multi-modal data aggregated as well as in time series.      &     G1-G3                   \\ \hline
\multirow{2}{*}{Relation \& Comparison}          & T3  & To assist comparing speeches in one speech level and between different speech levels.                                &  G1, G2, G4, G5                      \\
                                                 & T4  & To present the relations of multiple speeches derived by algorithm.               &   G4, G5                     \\ \hline
\multirow{2}{*}{Navigation}                      & T5  & To support browsing of speech videos guided by multi-modal emotional cues.    & G2, G3, G5                       \\
                                                 & T6  & To enable navigation of speech videos by emotional cues in video collections.    &  G2, G4, G5                      \\ \hline
Overview + Detail                                & T7  & To support understanding data between speeches, of entire speeches, and within a speech. &    G1-G5                    \\ \hline
Interactive Feature Specification                & T8  & To allow selection of specified speeches and factors.                                      & G1-G5                       \\ \hline
\multirow{2}{*}{Data Abstraction \& Aggregation} & T9  & To show the calculated correlation and distribution between effectiveness and factors.     &    G1, G2, G4, G5                    \\
                                                 & T10 & To provide effectiveness estimation on speeches in terms of factors.       &   G1, G3, G4, G5                     \\ \hline
\end{tabular}
\vspace{-0.5cm}
\end{table*}

According to the overall goals above, we derived 10 design tasks in \autoref{tab:tasks}, which are categorized as suggested by Kehrer and Hauser\cite{kehrer2013visual}. T1-2 focus on the visual data fusion, i.e., the way of presenting data to support further exploration. T3-4 focus on the relation and comparison between speeches, enabling users to understand the similarity and difference. T5-6 focus on the navigation of speeches, which provide users with interactions to explore in one speech or in a collection by emotional cues. T7 relates to the analysis from overview to detail. T8 focuses on the factor of interest to be dynamically specified by users. T9-10 mainly focus on the data abstraction and aggregation which support users to find out the hidden patterns or strategies in speeches and estimate the speech effectiveness 
using algorithms.

\section{Data and Analysis}

In this section, as illustrated in the upper part of \autoref{fig:systemoverview}, we describe the details of data collection (I), data pre-processing (II), factor calculation (III) and the method of effectiveness analysis (IV) based on the results of the  domain-centered design procedure. 

\subsection{Data}

There are three progressive steps for processing data into three levels: original data, features and factors. Each speech consists of: 1) the original video; 2) the scripts; 3) metadata such as the start and ending of the speech in the video; 4) information about the speech, including contest region, year, level and rank; 5) feature data extracted from the original video; 6) data of factors listed in \autoref{tab:pvalue}.

\subsubsection{Data Collection}

The entire database includes 203 videos from the World Championship of Public Speaking published online, including YouTube and WeChat channels. We collected the videos and metadata manually. The contest levels of the speech videos were recorded as a measurement of effectiveness: area, division, district, semi-final and final. The amount of videos for each level is approximately balanced, and we ensure that all collected videos are of good quality. Detailed information about our database is provided in Sect. 1 of the supplementary material.

\subsubsection{Data Pre-processing}

The part II of \autoref{fig:systemoverview} illustrates the data pre-processing step of our system. In order to acquire the previously mentioned factors, we extracted image frames, voice and text from the original video. The voice and text are aligned at the sentence level while the images remain at the frame level. 

\textbf{Facial emotional data:} We recognized discrete emotion types, valence and arousal of the speaker from the frames of video. Faces in frames and their positions are detected by face\_recognition\cite{face_recognition}. The faces are further clustered by DBSCAN\cite{10.5555/3001460.3001507} with the facial features extracted during detection to identify the faces of each speaker without the interference from others' faces in the video. AffectNet\cite{AffectNet} is a widely used database and baseline method for facial expression, valence and arousal computing in the wild. We used AffectNet to extract the facial emotion valence and arousal data. We extracted the facial emotion types using an emotion classification convolutional neural network by Arriaga et al. \cite{arriaga2017real}. 

\textbf{Textual emotional data:} We applied Azure Cognitive Speech to Text Service\footnote{\url{https://azure.microsoft.com/en-us/services/cognitive-services/speech-to-text/}} to convert spoken audio to script text with timestamps of each sentence and word. The method to extract the textual valence and arousal data came from the work of Wang et al.\cite{wang-etal-2016-dimensional}, which uses a regional CNN-LSTM model to analyze dimensional sentiment.
    
\textbf{Vocal emotional data:} With the timestamps of sentences, we split the audio into sentence-level clips and applied an open-source toolbox for multimodal emotion analysis developed by Buitelaar et al.\cite{8269329} to obtain  the vocal valence and arousal data of the clips.
    
\textbf{Non-emotional data:} We considered two non-emotional factors. The pauses between words and sentences as well as the words spoken per minute were calculated with the timestamps. The vocabulary level was measured with the Dale-Chall measure\cite{enwiki:1026007428} calculated with an open-source tool for assessing the readability level of a given text \cite{readability}.

\subsubsection{Factor Calculation} \label{section:factor calculation}

The raw data extracted from videos are time series of multi-modal valence and arousal data. As it is not intuitive for users to explore the data, we calculate some factors based on the raw data extracted in part III of \autoref{fig:systemoverview}. The calculation methods are as follows. The time series of multi-modal valence or arousal data is represented as $D=\{d_t^m\}_{t=1}^T$, where $d_t^m$ indicates the $t$-th sample of time series data in the modality of $m$. Similarly, the time series of multi-modal emotion type data is represented as $E = \{e_t^m\}_{t=1}^T$. The average factor represents the average value of a specific data modality over time:

\begin{equation}  \label{eq:1}  
	average = \frac{\sum_{t=1}^{T}d_t^m}{T}\\
\end{equation}

Volatility represents the change of data over time. We first normalize the data, and compute volatility according to Equations (\ref{eq:2})-(\ref{eq:3}).

\begin{equation} \label{eq:2}  
	D_{diff} = \{d^m_t-d^m_{t-1}\}_{t=2}^T\\
\end{equation}

\begin{equation}  \label{eq:3}  
    volatility = \sqrt{D_{diff}\cdot D_{diff}}\\
\end{equation}

Diversity represents the variety and relative abundance of the emotions\cite{quoidbach2014emodiversity}. We calculate it with Equation \ref{eq:4}. 

\begin{equation}   \label{eq:4}  
    diversity = \sum_{i=1}^{e}(r_i\times \ln{r_i})\\
\end{equation}

Here $e$ equals the total number of emotion types in $E$ and $r_i$ equals the proportion of $E$ that contains the $i$-th emotion.

In \cite{zeng2019emoco}, Zeng et al. explore the effect of emotion coherence across facial, text and audio modalities. Similarly, we calculate the value of arousal and valence coherence (defined in \autoref{eq:5}) where $std$ and $mean$ indicate functions of calculating the standard deviation and the mean value respectively. 

The superscripts represent the modalities of data, where w, v and f mean the textual, vocal and facial modalities.

\begin{equation}   \label{eq:5}  
    coherence = \frac{1}{T}{\sum_{t=1}^{T}\frac{std(d_{t}^{w},d_{t}^{v},d_{t}^{f})}{mean(d_{t}^{w},d_{t}^{v},d_{t}^{f})}}\\
\end{equation}

In interviews, experts estimated the last 20\% of a speech to be more important than other parts. So we calculated the final valence and arousal with Equation \ref{eq:6}.

\begin{equation}  \label{eq:6}  
	final = \frac{\sum_{t=0.8T}^{T}d_t^m}{T}\\
\end{equation}

For calculating final emotion, diversity and type ratio, we only select facial data as input, since textual data and vocal data are much sparser than facial data over time. For example, we could only extract an item of textual or vocal data from a complete sentence, while we can extract an item of facial data from each video frame. So for the same video, the amount of textual/vocal data is much sparser. We found processed textual/vocal results of these factors less convincing.

\subsection{Factor Effectiveness Analysis} \label{ordinal}

According to G1, we want to find out the relation between speech factors and effectiveness. Contest speech levels can be regarded as ordinal variables, whose relative ordering is significant. For example, the grades of 1-5 may respectively represent 5 levels ``area'', ``division'', ``district'', ``semi final'' and ``world final''. In the World Championship of Public Speaking, only those who rank at the top of a level will advance to a higher level. So we hypothesize that the higher the level, the more effective the speech is. Given the factors of a certain speech, the problem of predicting its level can be considered as an intermediate problem between regression and classification~\cite{gutierrez2015ordinal}. We first conducted the test of parallel lines and found the p-value is smaller than 0.05, proving that the level prediction problem is suitable to be solved by multi-class ordinal regression. Then we split this prediction problem into four sub-problems as shown in \autoref{fig:systemoverview} IV. In each sub-problem, we performed logistic regression on the odds ratio of each factor. Finally, we obtained the p-value of each factor in \autoref{tab:pvalue}, where we indicate factors calculated as significant relating to effectiveness. The result of our factor effectiveness analysis shows that the average of facial arousal, the average of vocal arousal and valence, the volatility of facial arousal and valence, the final facial arousal, and the ratio of facial happy expressions all have a significant correlation with speech effectiveness. Taking experts' advice into consideration, we selected typical factors and embedded them into our system. According to the result of the four sub-problems, we calculated the probability of the five levels as the factors change value. 

\section{System Overview}
\begin{figure}[t]
 \centering 
 \includegraphics[width=\columnwidth]{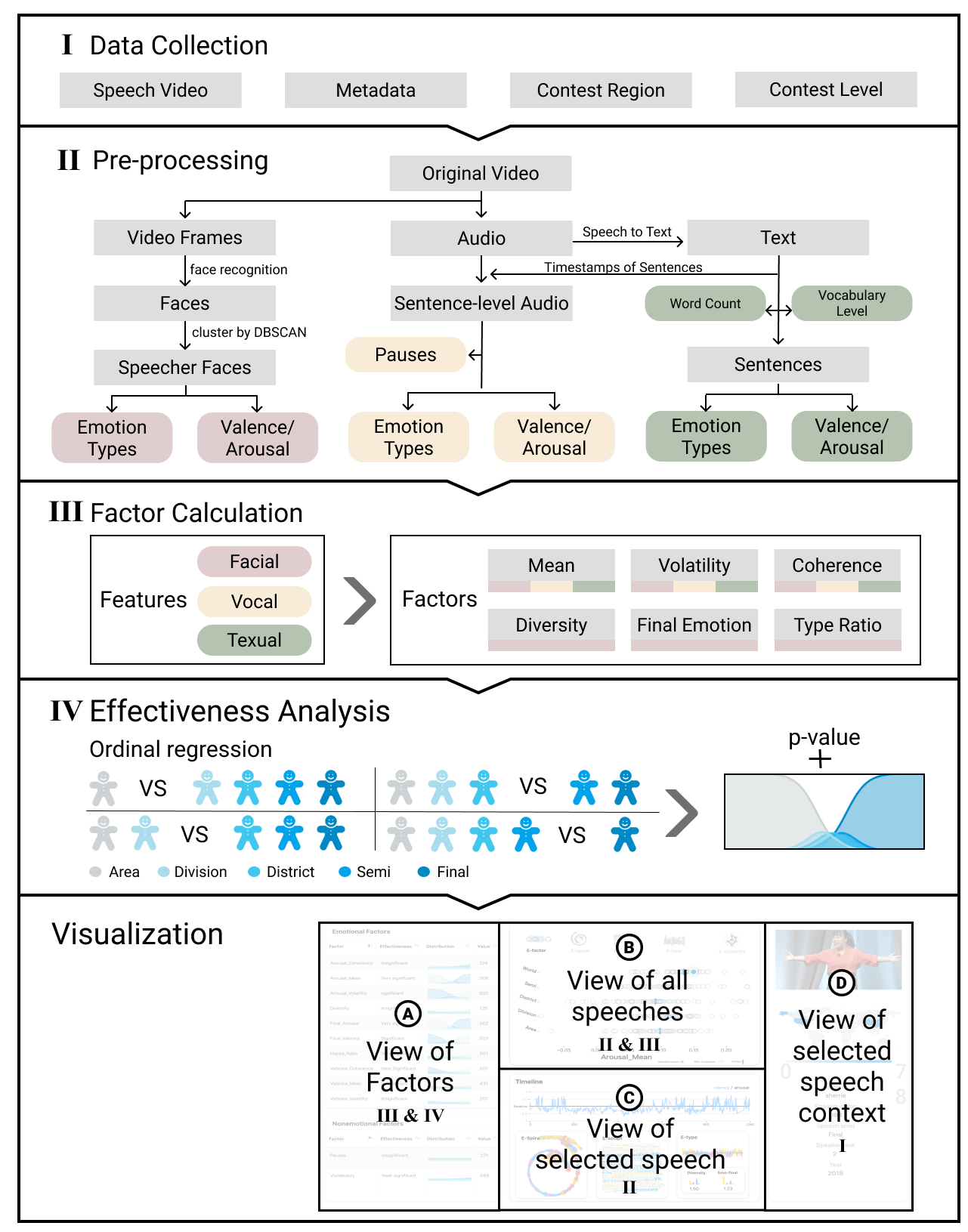}
 
\setlength{\belowcaptionskip}{-0.5cm}
 \caption{An overview of our analytic system.}
 \label{fig:systemoverview}
\end{figure}

 The design of the system was created in consideration of expert opinion, user feedback, and insight from the factor effectiveness analysis. \autoref{fig:systemoverview} illustrates the overview of our system architecture, which includes two major parts: a data \& analysis part and a visualization part. The visualization part adopts multi-modal and multi-level data from the first part, providing an interactive analytic interface for insights. 

\textbf{Views.} In the visualization part, as shown in \autoref{fig:MainInterface}, our system consists of four views: (A) the view of factors, (B) the view of all speeches, (C) the view of the selected speech and (D) the view of the selected speech context. 
In view A, the table displays the factors and p-values obtained from the factor calculation and effectiveness analysis steps. It helps users to understand the relation between speech effectiveness and various speech factors (G1), as well as the connection of emotional data to other speech factors. Given the range of factors of interest to the audience, a visualization system that provides an overview of all factors would be ideal for understanding their relation to effectiveness. 
The view B provides a panel to view all speeches for comparison, navigation and  exploration with five novel sub-views: E-factor, E-similarity, E-spiral, E-script and E-type (G1, G2, G4, G5); see \autoref{section:visualization} for details. We utilize the raw arousal and valence data from pre-processing phase and factor values from factor calculation phase to generate visualizations. These visualization techniques will be introduced in \autoref{section:visualization} in more detail. 
The view C contains four sub-views showing the data and visualizations of selected individual speech (G3). It helps users analyze a selected speech in more detail. 
The view D contains information about our database and detailed context information of the speaker (G2, G3). 
The four views of our system assist users to explore our database and find what factors affect the effectiveness of speech (G1, G5).

\textbf{Sub-views.} We mention that the view B and view C contain some sub-views above. The sub-views in view B are set to assist users to analyze overall trends, navigate and locate speeches of interest in our database. E-factor and E-similarity show the distribution of all speeches (G2). For E-spiral, E-script and E-type, we provide sub-views which aggregate the visualizations of all speeches (G5). The sub-views in view C are set to help users to observe the visualization of a selected individual speech in more detail using visualization tools such as E-spiral, E-script and E-type. We also visualize the original time series of valence and arousal data in the timeline sub-view (G3).

\textbf{Interaction.} We chose to design interactive steps of the system with Schneiderman's information seeking mantra as a guideline: ``overview first, zoom and filter, then details-on-demand." \cite{shneiderman2003eyes} The views are organized from left to right in terms of the level of abstraction of the data. We provide some interactions to support the overview-to-detail exploration process (G5). Upon selection of an effectiveness factor, E-factor will show the distribution of the factor values with all speeches. Users can also hover the mouse over the speeches to see the individual speech data and click to change the view of the selected speech. This interaction is supported by each of the sub-views in the view of all speeches (B). These sub-views aggregate all the visualizations of speeches and organize them by level. For deep exploration, users can click the sub-views in view (C) to generate a bigger visualization in a floating window. In E-similarity, upon clicking the dot representing an individual speech, a radar-like chart will be displayed in the right to show the predicted level of the critical factors of the selected speech (G4).

\section{Visualization Design} \label{section:visualization}

\begin{table*}[tb]
  \centering
  \caption{Visualization Methods and Corresponding Tasks.} 
  \label{tab:module}
  \begin{tabular}{|c|c|c|}
    \hline   
    Module & Description & Task\\
    \hline   
	E-factor & To evaluate hypotheses of interest about speech factors using the cumulative data of all speeches. & T1-T3, T6-T8  \\
    E-type & To understand discrete emotional data contained in emotional types, as well as their distribution over time. & T1, T3, T5, T7  \\
    E-script & To understand the emotion in speech scripts. & T1, T3, T6-T7  \\
    E-spiral & To provide an intuitive way of understanding the emotional shifts within speeches. & T1, T3, T6-T7\\
    E-similarity & To understand the similarity and the effectiveness estimation of speech factors in speeches. & T3, T4, T7, T10\\
    E-distribution & To understand distribution of factor effectiveness among speech levels. & T2, T3, T7-T9\\
    \hline
  \end{tabular}
\end{table*}

As with the system itself, our  visualizations were designed in iterations that began with interviews, and refined with user feedback.  In this section we present the final designs used in our system as well as the reasoning behind our visualization methods. The relation of our visualizations to our design tasks can be seen in \autoref{tab:module}.

We will introduce visualization techniques in two parts: visualizations generated from data of all speeches, including E-factor, E-similarity and E-distribution; visualizations generated from individual speech data, including E-spiral, E-script and E-type.

\subsection{Visualizations Generated from All Speech Data}

\subsubsection{E-Factor}

In our literature review of effectiveness in this inspirational speech contest, we found many claims by experts that particular speech factors have significant relationships to speech performance. This visualization aims to allow users to evaluate hypotheses of interest about speech factors at the macroscopic level of all speeches.

Upon selection of a factor in (A) the user is presented with many dots, with each dot positioned horizontally according to the cumulative amount of the factor in a speech (T2). The speeches are sorted vertically by the level of the speech.

As shown in \autoref{fig:MainInterface}(B1), the light blue rectangle for each level covers the middle 50\% distribution of the speeches  and the dark blue line indicates the median of each level (T3). A geographical analysis of factors is provided in view (D). By clicking a country on the map, the speakers belonging to the country will be highlighted, so users can analyze the regional difference between countries (T1).

\subsubsection{E-Similarity}

According to the experts' feedback, we found that they desired to compare the similarity between speeches (T4). To allow this comparison we chose  the five most significant factors as the speech's feature vectors, and used t-SNE\cite{van2008visualizing} to reduce the dimensionality of feature vectors to display all speeches on a two-dimensional map. The closer two speeches are to each other, the more similar the two speeches are. 

In order to better understand this relation, a radar chart displays the five most significant factors, and a given speech's estimated level based on the amount of each of the factors (T10). In \autoref{ordinal}, the result of 
ordinal regression contains the probability of the contest levels at a certain value of the factor. For a particular speech, we use the value of a certain factor to predict its level and use the radial coordinate of the radar chart to represent the predicted level. The larger the area of the polygon in the radar chart, the higher the predicted effectiveness of the speech is. As shown in \autoref{fig:MainInterface}  (B5), clicking a dot representing an individual speech in the scatter plot brings up a radar chart. Dots are color-encoded, allowing for rapid comparison of a speech's estimated level with its true value (T3).

\subsubsection{E-Distribution}
 We designed E-distribution in order to show how the effect of each factor changes the calculated probability of each level, which can be interpreted as a metric of effectiveness (T10). In \autoref{ordinal}, we obtained the probability of the five levels with respect to each factor. The five lines in the graph represent the distribution of probability of the five levels of the contest, with the same color encoding used as in E-Similarity. We can observe E-Distribution in \autoref{fig:MainInterface} (A). For example for the factor arousal mean we can observe larger values to the right of E-distribution result in higher probabilities of the darker line, or final level of the contest. 

\subsection{Visualizations Generated from Individual Speech Data}
\subsubsection{E-Spiral}

\begin{figure}[tb]
 \centering 
 \includegraphics[width=\columnwidth]{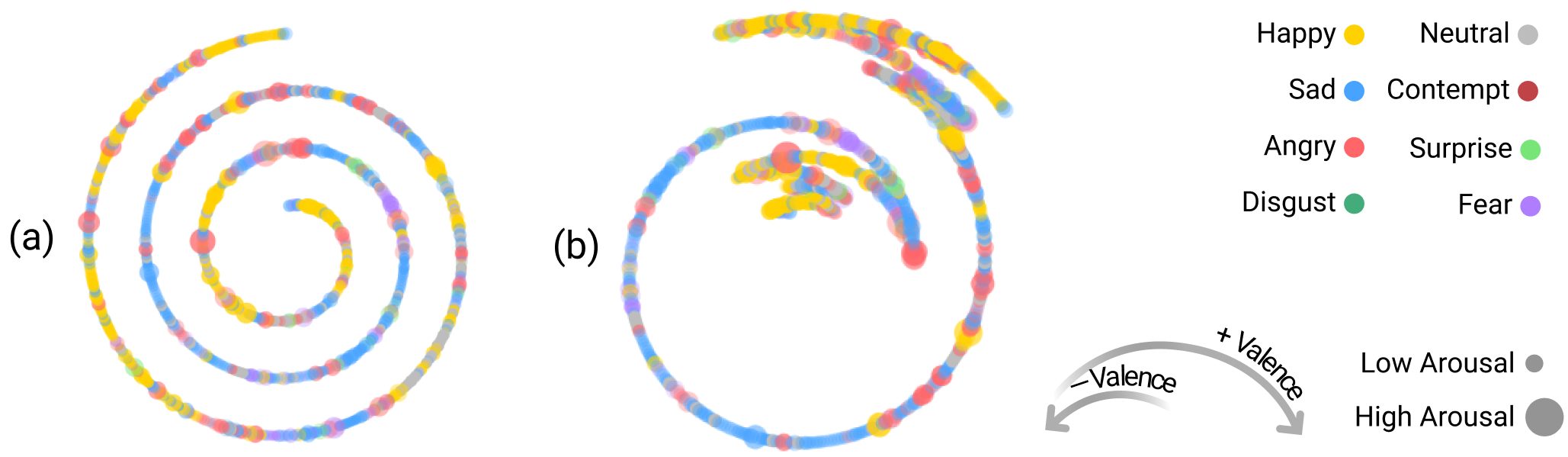}
 
\setlength{\belowcaptionskip}{-0.5cm}
 \caption{E-spiral: a novel spiral-type visualization of time-series emotion data. A comparison is given: (a) Spiral without turning points. (b) Spiral with turning points. Emotion shifts during speeches are intuitively presented with an interaction showing the detailed data.}
 \label{fig:Espiral}
\end{figure}

In our preliminary interview experts had suggested that emotional twists and turns may be important to consider in contest speeches. Preliminary data analysis early in our collection of speeches confirmed this hypothesis, as we found that there was significant statistical results for shifts in arousal (p-value 0.020) and valence (p-value 0.006). We sought to create a visualization that would present a compact view of these emotional shifts so that speeches could be rapidly compared. One option is spirals. According to a survey of timeline visualizations, spirals have advantages for displaying certain kinds of information, and their potential may not have been exhausted. Brehmer et al. found that spirals are ``appropriate for presenting many events within a single dense display." \cite{2017Timelines}. We therefore created a new form of spiral that shows the emotional twists and turns in a visually dramatic way. Clockwise and counterclockwise turns in the spiral indicate shifting negative and positive emotions, with sharp angles of the visualization showing the emotional turning points. Due to the compact structure, large scale comparison is possible, supporting comparison and navigation between speeches and within a speech (T7), as shown in \autoref{fig:MainInterface}(B2).

Based on the results extracted from the speech video, we identified the speaker's valence and arousal at regular intervals. Each circle appears in chronological order, starting at the center of the spiral. Significant shifts in the valence of speeches are reflected in the change of direction. As comparison of the emotional diversity of speeches was stated to be a priority in our pre-survey of experts, we further indicated the emotional type of the interval as the color of the circle. The circle radius represents the arousal of emotions and transparency represents the probability of the emotion being labeled correctly. 

E-Spiral is generated in polar coordinates with $\theta_n = \theta_{n-1} + 2\pi \Delta_{r}  p_{i}$. $\theta_n$ is the polar angle of the center of the $n$-th circle and $\Delta_{r} = r_n - r_{n-1}$ is the variation of the polar radius between the $n$-th circle and the $(n-1)$-th circle, which is a constant value since the spiral expands at a constant variation of radius.

The emotional turning points are generated based on the positive and negative changes of accumulated emotions in intervals. $E_i=\Sigma{a_n}$ is the accumulative emotion in an interval of 5 seconds, in which $a_n$ is one of the valence data in interval $i$. The spiral turns clockwise when $p=1$, while it turns counterclockwise when $p=-1$. The changing of $p$ decides the emotional turning points in spirals, which is calculated in  \autoref{eq:9}. The initial value of $p$ is defined by the emotion in the first interval, shown in \autoref{eq:8}. 

\begin{equation}  \label{eq:8}  
p_0=
\left\{
\begin{aligned}  
1 & , & E_0\geqslant0, \\  
-1 & , & E_0<0.
\end{aligned}  
\right. \end{equation}

\begin{equation}  \label{eq:9}  
p_{i\geqslant1}=
\left\{
\begin{aligned}  
-p_{i-1} & , & E_i * E_{i-1} < 0 \text{ and } | E_i - E_{i-1} | > 10, \\  
p_{i-1} & , & otherwise.
\end{aligned}  
\right. 
\end{equation}

With the help of E-spiral, as shown in \autoref{fig:Espiral}, we can clearly see the changes of emotion during the speech via the turning spiral. Interaction of rapidly skipping to the video frame of the selected speech by clicking the circle on the spiral supports rapid browsing with emotional cues (T5).

\subsubsection{E-Script}

\begin{figure}[tb]
 \centering 
 
 \includegraphics[width=\columnwidth]{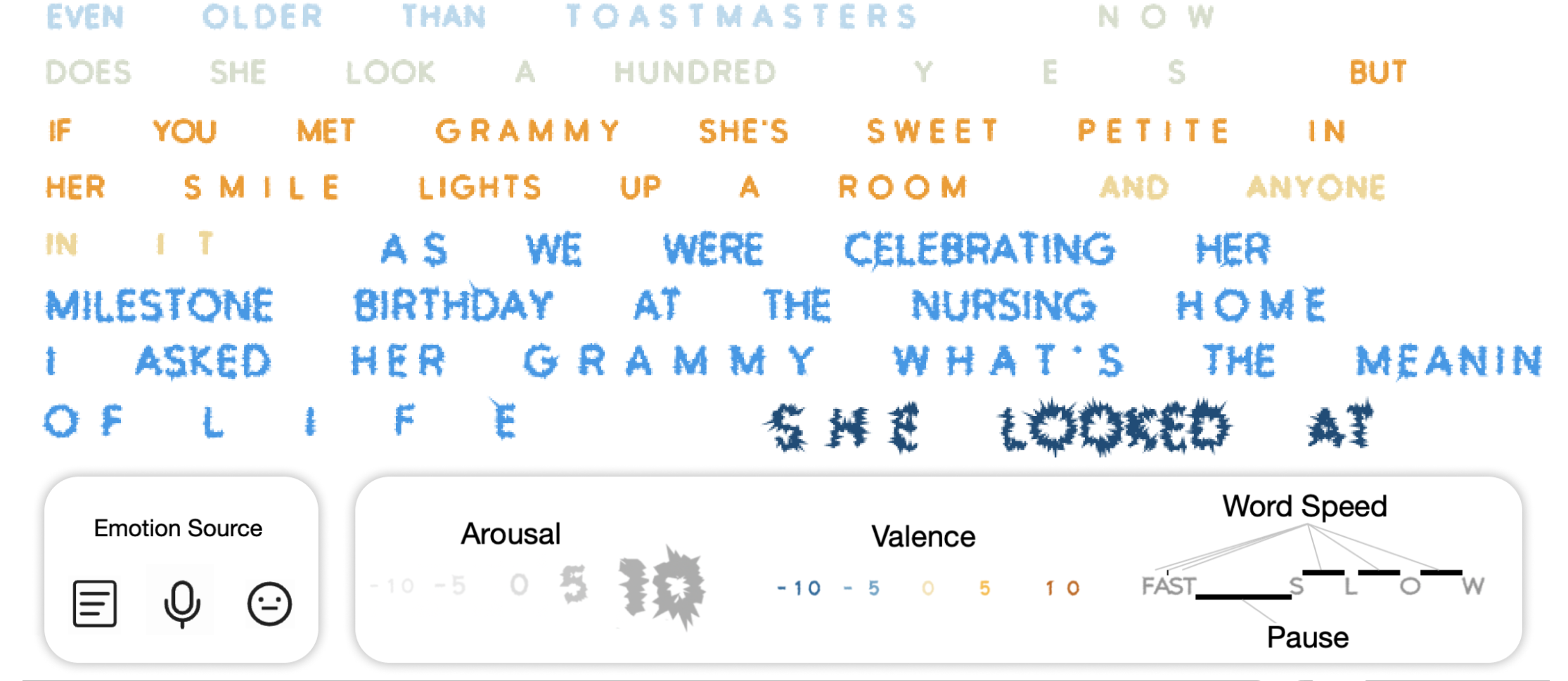}
 \setlength{\belowcaptionskip}{-0.8cm} 
 \caption{E-script: a novel visualization allowing key information about the text expression and emotional trends to be compared across whole speeches.}
 \label{fig:script}
\end{figure}
\begin{figure}[tb]
 \centering 
\end{figure}
How speech script content relates to key speech delivery information is the subject of much literature on speech giving. E-script allows fine grained understanding of how  multi-modal speech emotion (T5), word speed, and pauses relate to the timing of each word of a speech (T1).

As experts advised that effective use of pauses and word speed was important to the emotional delivery of  script content, we sought to indicate these factors in an intuitive way. We indicated the word speed by the tracking between letters in a word and pauses as the spaces between words.

In a novel approach, E-script aims to provide an audience with ordinary visualization literacy with an intuitive way of understanding script emotional data by changing letter shape to encode quantitative information (quantitative glyphs).  E-script highlights emotionally intense, high arousal moments. Sievers et al.\cite{2020A} found that the shape of lines can visually encode emotion and is closely related to arousal, and can be seen as independent from valence.  We supposed that this visual mapping might be applied in visualizations with line shape. We implemented this visual mapping of arousal directly in the speech script by connecting the line based javascript font Leon Sans\cite{leonsans} with the arousal information. The size of the text was also changed in order to highlight emotionally intense moments in the speech. The method of mapping of the color to the valence and arousal in the speech script can be seen in the work \cite{2020EmotionMap}. 

No papers have been found in visual analytics that involve letter shape adjustment in order to convey quantitative emotion. Brath et al. \cite{BRATH201659} surveys existing text visualizations, noting only one case of quantitative glyphs, used for indicating prosody. The use for intuitive communication of emotion here can be seen as a novel expansion.

\subsubsection{E-Type}

\begin{figure}[t]
 \centering 
 \includegraphics[width=\columnwidth]{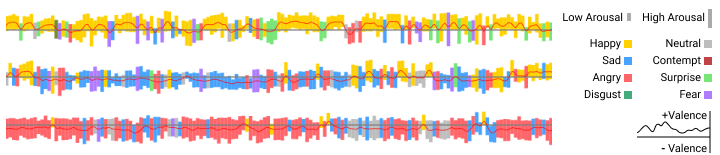}

 \caption{E-type: A linear visualization of discrete emotional type together with continuous emotional data.}
 \label{fig:Emo-type}
\end{figure}
As discussed in the interview section, an open question for speech experts is the role of the categories of emotions in speeches, as well as the temporal distribution of emotions (T1). E-type allows continuous valence and arousal data to be compared to discrete emotional type data across time in speeches. We supposed that a linear visualization might be more intuitively understandable to non-expert audiences. 

As shown in \autoref{fig:Emo-type}, we use a series of rectangles to show the categories of emotions and sample the emotional data of each speech evenly 200 times to get 200 rectangles. The color of rectangles represents the category of emotion. The height and $y$-coordinates of the rectangle represent the arousal value and valence value respectively. We use a red line to connect the center points of the rectangles, which indicates the change of valence. With E-Type, users can grasp the proportion of each category of emotion more clearly. In contrast to E-spiral, E-type offers a more fine tuned view of continuous emotional data.

\section{Evaluation}

Our evaluation focuses on two main goals as categorized by Lam et al.\cite{lam2011empirical}: user experience (UE), which we address with a usability test; visual data analysis and reasoning (VDAR), which we address with a user survey and a case study.

\begin{figure}[t]
 \centering 
 \includegraphics[width=\columnwidth]{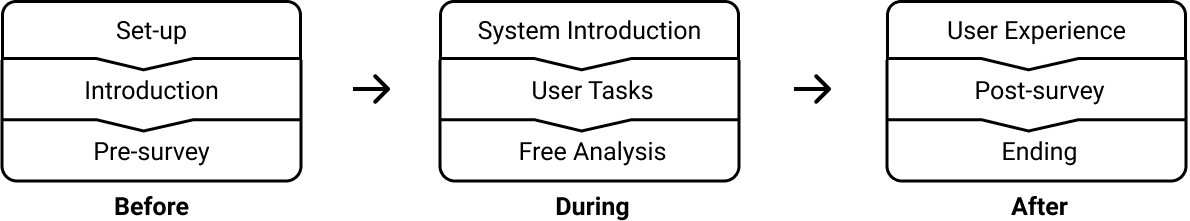}
\setlength{\belowcaptionskip}{-0.5cm}
 \caption{Our evaluation procedure.} \label{fig:evaluation}

\end{figure}

\subsection{Study Design}

The procedure of our evaluation is shown in \autoref{fig:evaluation}. Before participants experienced E-ffective, we introduced the project and related terms, as well as pre-surveyed their opinions of speech factors (VDAR). Next, they were introduced to the system, instructed to complete a series of user tasks,  then freely analyzed speeches by using the system. Finally, they finished a user experience interview (UE) and a post-survey (VDAR).

\textbf{User Tasks:} The user tasks focused on instructing the participants to better understand the main function of the visual analysis methods and the whole system. We designed 10 user tasks aiming at both visualization design (UT1-UT6) and system design (UT7-UT10). The detailed tasks are listed in Sect. 4.2 of the supplementary material. 
Participants were limited to using specified visualizations to complete UT1-UT6 while they were free to use any visual analysis methods to complete UT7-UT10.

\textbf{Free Analysis:} 15 minutes of the evaluation was dedicated for participants to analyze speeches by using the whole system in a self-directed manner. During the analysis, participants were encouraged to freely speak out their intent in using the system as well as their findings on speech effectiveness. The purpose of the free analysis procedure was to observe the creative ways users reasoned using the system and record meaningful insights made by participants.

\subsection{Participants}

Our system is designed for both experts and novices. We recruited 16 participants for evaluation, including 8 experts in public speaking and 8 novices. All participants had competed in the World Championship of Public Speaking, and all were active in the field of public speaking. None of the participants had expert level visualization literacy.

The experts participants (E1-E8) all had rich experience in public speaking. Almost all of them (7 of 8) have participated in over five speech contests and five of them have trained others for public speaking as an occupation for more than five years. Half had STEM educational backgrounds. Five of them had watched more than fifty speech contests. It is worth mentioning that one of the volunteers once won the second place in the world final of the contest, and another was a two-time world semifinal competitor.

Of the novice participants (N1-N8), seven of them had competed in the contest at least twice. Half of them had STEM educational backgrounds, the same ratio as experts. Most novices had watched more than ten speeches, and they all desired to improve their speech skills by learning from excellent speakers.

Detailed information about the participants is provided in Sect. 4.1 of the supplementary material.

\subsection{Usability Testing}

In our usability test, we wanted to evaluate how useful and easy to use our system and visualization methods are.

\begin{figure}[tb]
 \centering 
 \includegraphics[width=\columnwidth]{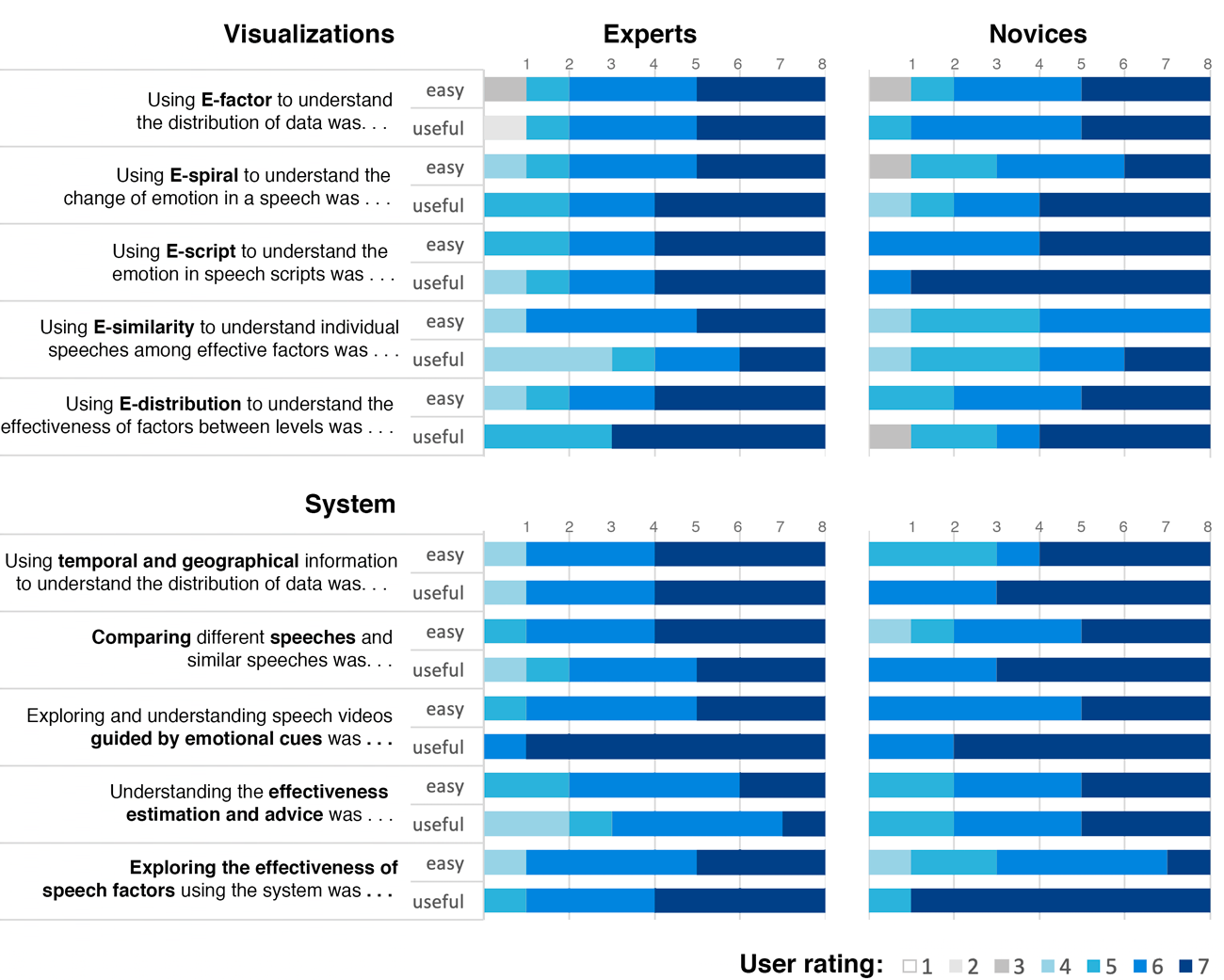}
 
\setlength{\belowcaptionskip}{-0cm}
 \caption{Result of the user experience questionnaire.}
 \label{fig:userexperience}
\end{figure}

Following the completion of both the user tasks and free analysis, user experiences of participants were scored in 7-point likert scale for usefulness and ease of use in a user experience questionnaire. As shown in \autoref{fig:userexperience}, the questionnaire evaluates both visualization design and system design.

\begin{table}[]\small
    \caption{Statistical Data about the Results of the Questionnaire.}
    \label{table:question} 
    \centering
\begin{tabular}{cccclcc}
\hline
                         & \multirow{2}{*}{Question} & \multicolumn{2}{c}{Visualization Design} &  & \multicolumn{2}{c}{System Design} \\ \cline{3-4} \cline{6-7} 
                         &                           & Mean                & Std                &  & Mean            & Std             \\ \hline
\multirow{2}{*}{Experts} & Usefulness                & 5.950               & 1.218              &  & 6.200           & 0.966           \\
                         & Ease of Use               & 6.075               & 1.047              &  & 6.200           & 0.822           \\ \hline
\multirow{2}{*}{Novices} & Usefulness                & 6.150               & 1.051              &  & 6.575           & 0.636           \\
                         & Ease of Use               & 5.900               & 1.033              &  & 6.050           & 0.876           \\ \hline
\end{tabular}
\end{table}

The description and results of the user experience questionnaire are provided in \autoref{fig:userexperience}. The mean and standard deviation of the result are shown in \autoref{table:question}. 
From the results, we can conclude that the mean scores are all higher than 5.9, showing that our system and visualizations are useful and easy to use for experts and novices.
The standard deviations of the visualization design questions are higher than the system design questions. Given the relatively low visualization literacy of the participants, perhaps these novel visualizations may be relatively complex, leading to different levels of understanding of them.

In \autoref{fig:userexperience}, we can see that both experts and novices thought that E-similarity is less useful and less easy to use than the other visualizations. There was additional feedback from several participants  that the five significant factors used in this visualization were not comprehensive enough to understand a speech. 
Also, several participants had difficulty conceptually understanding the t-SNE layout, leading to further difficulty in understanding why two speeches appeared similar.
Experts preferred E-distribution, possibly because E-distribution can help them intuitively find out how effectiveness changes in regards to factors that they are more familiar with. Novices preferred E-script, and several of them said that they thought it was very useful to explore the emotion shown inside the original speech script. Many found the pauses and word speed information that was missing in other visualizations as useful. 
For questions about system design, participants reported more satisfaction with the guidance of emotional cues. They agreed that emotion plays a key role in inspirational speeches. Novices reflected that our system can help explore the effectiveness of speech factors. They gave the system higher scores than the experts in terms of usefulness, perhaps because experts have more experience about speech effectiveness than novices, and thus they have their own opinions that may differ from our system results. Novices scored the system lower in terms of the ease of use than experts, perhaps because they lack experience in public speaking and the terms associated with competition success. On the whole, they needed more time to understand and grasp how to use the system.

\subsection{User Survey} \label{sec:user survey}

\begin{figure}[t]
 \centering 
 \includegraphics[width=\columnwidth]{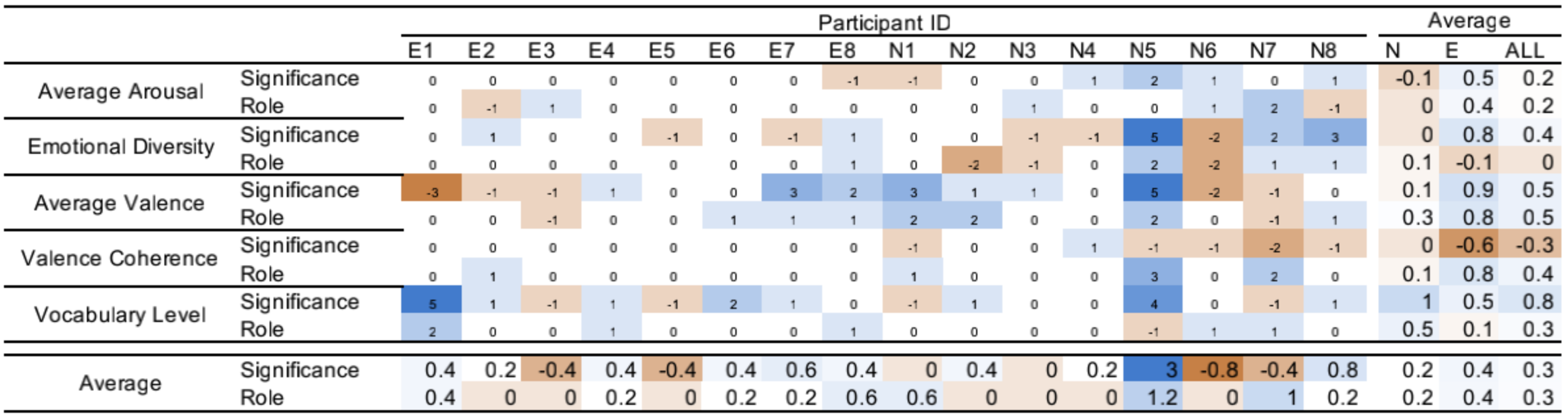}
 
\setlength{\belowcaptionskip}{-0.5cm}
 \caption{Result of the user survey. Comparing the opinions before and after using the system. Darker blue values
 mean the opinion of a user changed to be closer to the system suggestion, and brown means the opposite.}
 \label{fig:prepost}
\end{figure}

Speech experts and novices were interviewed before using the system to determine their views about the importance of factors in the contest. The participants were asked to give their opinions on whether they thought a factor would significantly impact performance between contestants, and then to predict the typical role the factor had on performance. 

Afterwards, they were surveyed again to see whether their opinions on the surveyed questions had changed. The participants generated their hypothesis on the speech factors when they were pre-surveyed. They further examined their hypothesis by using the system. In the post-survey, we collected their results , which we later used to understand their hypothesis generation and reasoning.

\textbf{Surveyed factors and measurement.} The factors included 1 factor (average arousal) determined by our analysis in \autoref{ordinal} to have significant influence, 2 factors (valence coherence, vocabulary level) to have near significant influence, as well as 2 factors (emotional diversity, average valence) to have insignificant influence. The significance is measured in 1-7 from very insignificant to very significant, and the relationship is measured in -3-3 from very negative to very positive.

\textbf{Results processing.} We adapt our analysis results as ground truth for judging the directions of opinion changes, so that the judgment of whether the participants were influenced positively or negatively can be concluded. The results of pre-survey ($R_{pre}$) and post-survey ($R_{post}$) were further processed. For significance results, the final score of each question is $F_{s}=R_{post}-R_{pre}$ if the suggested significance in our factor analysis is significant and $F_{s}=R_{pre}-R_{post}$ if insignificant. For relationship results, the final score of each insignificant factor is $F_{r}=\vert R_{pre}\vert - \vert R_{post} \vert$ to measure the change towards 0. The final score of each significant factor is $F_{r}=R_{post}-R_{pre}$ if the relation is positive and $F_{r}=R_{pre}-R_{post}$ if the relation is negative. The processed result is shown in \autoref{fig:prepost} as a heat map.

\textbf{Survey prediction.} The results of user survey prove that our system influenced the participants' opinions after using the system. In general, we can see that our system had a positive impact on participants' understanding of speech factors. 
Changes in the opinions of experts show that they are slightly more affected than the novices. But we can also find that, more experts insisted on their own opinion on factors than novices. 
In the respect to individual factors, there is a bigger change in the opinion on the significance of vocabulary for both experts and novices. However, in terms of all factors, there is a difference between the significance and relationship in the changes of participants' opinions. The impact of the system on factor significance seems to have more negative results than that of factor relationship. 
The reason may be that in the pre-survey participants expected the significance too high, but from the system they the factors less significant than they thought. So that in the post-survey they overcompensated and rated the factor as under significant. Overall, the results of user survey indicate that our system enables users to update their opinions based on the findings by exploring speech factor effectiveness.

\subsection{Case Study}

1) In our evaluation sessions we often found experts identifying effective and ineffective contest strategies (G5). Upon finding our system had showed a trend among speakers at the lower levels having large ratios of happy expressions, a past world second place winner of the contest commented: ``In my experience there are two kinds of speakers at the lower levels: one shows a lot of happiness and jokes. Another kind just tells a sad story. The emotion doesn't change so much." Later this participant viewed the speeches sorted by emotional diversity with E-factor and clicked on a speaker with very low emotional diversity. The participant was surprised to find it was the last world champion. This countered her earlier pre-survey opinion that diversity is very significant and high diversity is better. She then carefully verified the accuracy of the facial data using E-spiral by mousing over dots on the visualization to reveal facial expressions. She then reasoned that online speeches may not need as much emotional diversity, ``this is an online speech, it is a little bit different from a speech in a big meeting room. When I was on the stage, I was facing a lot of people and they really couldn't see my face, they had to look at the big screens." This new hypothesis revealed a limitation of our system: we had not separately labeled and processed the results of online and offline speeches. In other interviews additional tags were suggested regarding yearly trends,  gender, and other factors.

Exploring the context of different speaking strategies was often focus of experts during the evaluations.

One expert with contest experience in both China and at the world semifinal level in the United States explored the geographical differences among competitors (G2). He found the vocabulary of contestants in China to be lower and after exploration of the difference in China and the USA, he sorted contestants by vocabulary with E-factor to find a world finalist contestant. He then used E-script to find the many difficult words used by the competitor: ``If he is competing in China, he is going to lose the competition.
Native speakers tend to have high levels of English vocabulary, but when they are addressing to different groups of audiences they probably should use a lower level of vocabulary." This new hypothesis countered his pre-survey prediction that winning speeches were more likely to have higher vocabularies.

Expert opinion about the use of vocabulary differed strongly in our interviews, as well as among existing literature. A previous world champion supported his view that the best speeches have simpler vocabularies with the claim ``winning (world champion) speeches over the preceding 17 years ranged at grade levels between 3.5 and 7.7" \cite{donovan2014speaker} What this claim does not show is that in our survey, speeches at the highest level of the contest have larger vocabularies on average than any other level. 

2) Several users and experts found additional applications for E-script to understand critical factors of speech delivery. One novice and former district contestant found E-script was intuitive to understand: ``(E-script is) useful for me because if I can see where I need to slow down my pace. And I can also see where the color matches the emotion and if it fits." The application of E-script by other users to evaluate emotional coherence across modalities as well as paces and pauses was observed in our evaluation.

Furthermore, an expert who works full time in training for public speaking gave E-script full marks for usability and easiness, and provided suggestions for development: "I love it, I can see how people can use pauses to make their speech more powerful. In the future there could be a function where people could view and collaborate directly on the use of pauses in the script." 

\section{Discussion}

In this section, the limitations and future directions of our system are elaborated on.

\subsection{Data and Processing}

While collecting speech videos of over 200 inspirational videos from open access channels, we tried to keep a balance of quality and quantity.  The size of the database is still insufficient for 
conclusive evidence about the factors we studied.  Therefore, we plan to enlarge the size of the database while further applying the system to other practical domain applications. 

\subsection{Factors and Analysis}
In this paper we focused on the major factors taken from our literature review and interviews of domain experts. However, determining what factors affect the performances of speeches still remains a complicated problem. It is hard to include all the factors that matter and reduction is necessary for better quantitative analysis methods. 

In order to better find out the factors affecting the effectiveness of the speech, we extracted some factors from the valence and arousal data, instead of directly analyzing the valence and arousal data. While this allows users to more intuitively explore what factors affect the effectiveness of the speech, on the other hand it may be oversimplified. Current effectiveness analysis on factors is limited to the univariate linear regression analysis, and the system does not consider the interaction between variables and other complex relations.

\subsection{Limitations in Evaluation}
In our evaluation we aimed to assess user experience and visual data analysis and reasoning. There are limitations to the results we obtained, especially in the case of the user survey. While we compare results before and after use of E-ffective to the estimated results of our model, these results cannot be construed to suggest learning. Problems in the study design may influence the results, including the small sample size. Additionally, factors such as the reputation of the developers of the system or the evaluation organizer may influence the credibility of our prediction outcomes. The difference of the reliability of our models on one type of data may also influence the perceived reliability of other types of data. The varying accuracy of the various models we use in our system are likely to skew the results of our post-survey.

\subsection{Limitations in Domain}

The effectiveness of a speech is a very subjective issue, and there is no clear and quantifiable evaluation standard. Different people may have very different opinions of the same speech, which depends on the preferences of the audience, cultural differences, and many other factors. We try our best to analyze the patterns of speeches in an objective, data-driven way. In addition, emotion does not play a key role in all types of speeches. Many public speaking experts consider emotion to play a special importance in inspirational speeches. The results of our current visual analysis are not applicable to all situations.

\subsection{Generalization}

The E-ffective system proposed in this paper focuses on exploring the effectiveness of factors in inspirational speeches. Through the evaluation part of our work, it is proved to be useful, easy-to-use, and fits the domain requirements. Moreover, insights were made by users that had different levels of experience in the domain. However, the potential of our system is not restricted to the domain of inspirational speeches. We can see the possibilities of extending the system to analyze the effectiveness of factors in other kinds of speeches.

\section{Conclusion}
In this paper, we propose E-ffective, a visual analytic system built for speech experts and novices to evaluate the effectiveness of speaking strategies. Our system primarily addresses factors involving emotional expression in an inspirational speech contest. Our evaluation studies confirmed the utility by a usability testing study, and the ability of participants to analyze and reason using the system was demonstrated in the case study and two surveys. In order to support the needs of our users we found many potential factors influencing effectiveness in the competition. From algorithms and visualization methods we found what factors were tied to effectiveness. The importance and utility of these factors were later verified in our evaluation. Two novel forms of visualization, namely E-spiral and E-script were developed to further assist users to understand critical factors and their application in speeches. 

In future work, we have already begun to expand our database of inspirational contest speeches as well as expand our methods to create other kinds of speech  effectiveness databases. We also plan to improve our analysis methods by means of considering the interrelation of factors and further expansion of the considered factors. Finally, we see the potential for additional visualization methods to be developed to more intuitively display factors critical for the effectiveness of speeches. 

\acknowledgments{This work was supported by the Beijing Natural Science Foundation (4212029), the Natural Science Foundation of China (61872346, 61725204), Alibaba Group through the Alibaba Innovative Research Program and the 2019 Newton Prize  China Award (NP2PB/100047).}
\bibliographystyle{abbrv-doi}

\bibliography{E-ffective}
\end{document}